\documentstyle[preprint,aps]{revtex}

\begin{document}
\draft
\preprint{RU-98-21}  
\title{Correlation between Compact Radio Quasars\\ and Ultra-High Energy Cosmic
Rays\\
}
\author{Glennys R. Farrar}
\address{Department of Physics and Astronomy\\
Rutgers University, Piscataway, NJ 08855-0849, USA\\
}
\author{Peter L. Biermann}
\address{Max Planck Institut f{\"u}r Radioastronomie\\
Auf dem H{\"u}gel 69, D-53121 Bonn, Germany\\
}
\date{\today}
\maketitle
\begin{abstract}
 
Some proposals to account for the highest energy cosmic rays predict that
they should point to their sources.  We study the five highest energy
events ($E>10^{20}$ eV) and find they are all aligned with compact,
radio-loud quasars.  The probability that these alignments are
coincidental is 0.005, given the accuracy of the position measurements
and the rarity of such sources.  The source quasars have redshifts
between 0.3 and 2.2.  If the correlation pointed out here is confirmed
by further data, the primary must be a new hadron or one produced by a
novel mechanism. 

\end{abstract}
\pacs{
{\tt$\backslash$\string pacs\{\}} }

\narrowtext

The nature and origin of the highest energy cosmic rays ($E \ge 10^{20}$
eV) is one of the major questions in physics and astronomy.  Energies up
to $3.2 \times 10^{20}$ eV\cite{flyseye}, corresponding to center-of-mass
energies up to $\sqrt{s} \approx 800$ TeV, have been observed for the
primary interaction with an atmospheric nucleon.  The showers produced
by these cosmic rays indicate that the primary is a hadron such as a
proton or light nucleus\cite{flyseye,akeno:cluster,halzen}, although a
photon is not completely excluded.  Astrophysical mechanisms to
accelerate protons to energies of up to $10^{21-22}$ eV have been
identified\cite{pb:1021}, but they require exceptional sites.  In his
pioneering analysis, Hillas\cite{hillas} observed that the source could be
a radio galaxy or quasar, and not much else, based on general
considerations.   

The conundrum is that nucleons, nuclei, and photons of energy greater
than about $5 \times 10^{19}$ eV  have a non-negligible scattering
cross section from the cosmic background radiation (CBR), causing
their energy to be reduced to this level if they travel far enough
through the CBR.  This is known as the Greisen, Zatsepin, Kuzmin (GZK)
limit\cite{greisenzatsepin}.  If a proton or photon arrives at Earth
with an energy greater than $10^{20}$ eV, it is exceedingly unlikely to
have originated further than 50 Mpc\cite{elbert_sommers,f:114}, whereas
suitable astrophysical acceleration sites are located at greater
distances\cite{biermann:rev}, and indeed none are found within the
expected scattering cone of the higest energy event at less than the
GZK distance\cite{elbert_sommers}.   A heavy nucleus might survive the
requisite distance\cite{stecker} and would have a larger scattering
cone, but it is difficult to reconcile the observed flux with the $\le
10^{-5}$ Fe to p ratio corresponding to cosmic abundances.

Proposals to resolve the puzzle range from positing superheavy relics
-- topological defects or heavy particles -- whose decay produces  
nucleons and photons within the GZK
distance\cite{topo}, to positing new particles or mechanisms which evade
the GZK bound\cite{f:114,f:104,weiler-nu}. 

We study here a prediction of the latter scenarios: each UHECR
should point directly to its source, which can be at cosmological
distances.  This was already remarked for the case the UHECR primary
is a new, neutral, stable or very long lived supersymmetric hadron of mass
a few GeV\cite{f:114}.  It also applies to the mechanism of ref.
\cite{weiler-nu} in which neutrinos of energy $E_\nu = M_Z^2/(2 m_\nu) =
4 \times 10^{21} ({\rm eV}/m_\nu)$ annihilate with dark matter neutrinos
via $\nu \bar{\nu} \rightarrow Z^0 \rightarrow q \bar{q}$ to produce
hadronic jets.  This is because the opening angle between the incident
neutrino and a particle of energy $E$ produced in the $Z^0$ decay is $\delta
\theta \le M_Z/(2 E) \sim 10^{-9}$.  Furthermore 
in both these examples the primary has negligible energy loss except for
redshift, so can be produced in distant objects such as QSOs.

By contrast, a proton or nucleus would neither point to an astrophysical
source nor be associated with a large $z$ QSO, since its scattering from
the CBR excessively dissipates its energy unless $z < 0.01$.  The rms
deflection of a proton of energy $E$ traversing randomly-oriented patches
of magnetic field having rms value $\delta B$ and scale length $\lambda$
is given by      
\begin{equation}
\label{angdev}
\delta \theta \sim 7.2^o \sqrt{d \lambda / 200 ({\rm Mpc})^2} (\delta
B/E) (100 {\rm EeV}/10^{-9} {\rm G}),
\end{equation}
where $d$ is the distance from the source to the observation 
point\cite{akeno:cluster}.  Note that there is some evidence for sheets
of much higher magnetic field\cite{pb:magfield}.

We list in Table \ref{tab:events} all the UHECR events whose energy is
at least 1 $\sigma$ above $8 \times 10^{19}$ eV and whose direction is
known with a solid angle resolution of $10~ {\rm deg}^2$ or better.
The energy cut is imposed in order to exclude contamination from events
which may be due to proton primaries and the angular resolution 
requirement is necessary to reduce random background.  The angular
resolution in general improves with energy, so both cuts would have to
be relaxed in order to enlarge the sample.

A few comments on Table \ref{tab:events} are in order.  The error bars on
the energy of the Fly's Eye event include systematic as well as
statistical uncertainty.  The parameters of the Haverah Park events are taken
from \cite{watson:HEPiN}, with errors on the positions determined by us using
the formulae given in ref. \cite{hollows}.  The angular error in the
longitudinal direction, relevant for $\Delta \Omega$, is $\Delta
\alpha * cos \delta$ where $\Delta \alpha$ is the error in right
ascension and $\delta$ is the declination.  AGASA reports an angular
cone radius, denoted $\sigma_r$ below, defined such that 68\% of the
events would be contained within the error cone: $1^o$ from
statistical error alone and $1.6^o$ including systematic
errors\cite{agasa:210,akeno:cluster}. Our information on Ag110 come 
from ref.  \cite{akeno:cluster}, which does not give an error on the
energy measurement, although ref. \cite{agasa:res} quotes a 30\% error
in general so we expect this event satisfies our cut.  The high energy
event observed by Yakutsk, with $E=1.1 \pm 0.4 \times 10^{20}$, has
too large an angular uncertainty and is not $1 \sigma$ above $ 8 \times
10^{19}$ eV so it is not used. See \cite{watson:HEPiN} and
\cite{akeno:cluster} for other high energy events which we cannot use.  

A correlation with quasars has already been noted for the two highest
energy events.  Elbert and Sommers\cite{elbert_sommers} searched
within $10^o$ of the highest energy event, the $320^{+92}_{-94}$ EeV
event observed by the Fly's Eye group\cite{flyseye}. They identified
the exceptionally radio-loud quasar 3C 147 as an ideal source, aside
from its  extreme distance.  With a redshift of 0.545, its distance is
of order 2000 Mpc.  Biermann\cite{pb:StockholmCR} pointed out that
another remarkable quasar, PG0117+213, is inside the error cone of the
second highest energy event (210 EeV)\cite{agasa:210}.  At $z=1.493$,
about 3500 Mpc, the distance also seemed too great for it to be seriously
considered as a source.   

The surface density of QSO's is large enough that these two alignments
are not statistically significant and may be accidental.  However
acceleration of protons to $ \ge 10^{21}$ eV requires a remarkable
source, so if the hypothesis is correct it may be possible to
identify a more restricted class of sources, with low surface density,
for which the correlation is statistically significant.  

One of the best-motivated cosmic ray acceleration regions is the jet of an
AGN (Active Galactic Nucleus), where relativistic shocks and large magnetic
fields are found.  Depending on the age of the AGN, the orientation of its
jet with respect to Earth, the ``clouds" surrounding the inner accretion
disk and their relationship to the jets and Earth, and the amount of dust
in the host galaxy, the same source can be a blazar, radio galaxy, or a
quasar.  It can be unusually bright at visible wavelengths and/or optically
variable.  The shape of its radio spectrum depends on whether it is a
full-sized quasar or compact.

One would like to impose the seemingly trivial criterion that the energy
flux in cosmic rays implied by the UHECR observation itself, not be much
larger than the total electromagnetic energy output of the source.  However
even this simple condition is not straightforward to implement for AGNs,
due to their directional anisotropy.  For instance a blazar pointed away
from us has a much higher total energy output than evidenced by its
observed luminosity.  Moreover the energy output in a given wavelength band
can differ by orders of magnitude depending on the intervening material
which can ``reprocess" the electromagnetic energy.

In examining the properties of 3C 147 we noticed that it is a
compact quasar -- that is, its jets are only about 1/10 the size of a
full-sized quasar with radio lobes.  Other indicators of its compact
character are its optical variability and the fact that its spectrum
is cutoff at low radio frequencies\cite{odea:98}.  An anomalous spectrum
such as this is characteristic of compact radio-loud sources (CSS and
GPS)\cite{odea:98}.  We therefore defined the following specific
criteria for compact QSOs (CQSOs):  
\begin{itemize}
\item {\it QSO in the NASA/IPAC Extragalactic Database (NED).}
\item {\it Radio-loud.} In practice, we required that the object appear
in the Bonn catalog\cite{bonncat}. This is a compilation containing
1835 radio sources including all those whose flux density is $\ge 1$
Jy at 5 GHz.  The whole sky, excluding the galactic plane ($|b^{\rm
II}| < 10^o$), is covered.  The surface density of this class of
sources is therefore $1835/(34100~ {\rm deg}^2) = 0.054~ {\rm
deg}^{-2}$. 
\item {\it Flat or falling radio spectrum at low frequencies.} One-third
of the Bonn catalog entries have a flat or falling spectrum at low
frequencies, so the background surface density of the CQSO category 
is 0.018 deg$^{-2}$.
\end{itemize}

We first determine the probability that the UHECR events actually
point directly to the candidate sources, given the experimental
measurement errors.  After that we find the probability that randomly
distributed compact QSO's, given their surface density, would have an
equally good alignment to that observed.

We employ the method of maximum likelihood, which is a standard
tool in High Energy Physics.  For a concise review see the probability
and statistics sections of ref. \cite{pdg96}.  One makes use of the
quantity 
\begin{equation} 
\label{chi2}
\chi^2 \equiv \Sigma_{i=1}^{N_U} \{
|x_i - x_i^0|^2/\sigma^2_{x,i} + |y_i - y_i^0|^2/\sigma^2_{y,i} \}
\end{equation}
where $N_U$ is the total number of UHECR events in the analysis,
$\sigma_{xi}$ is the error on the $x_i$th coordinate, $x_i$
is the measured value of the coordinate (the UHECR position) and $x_i^0$
is the (hypothetically) true value of the coordinate, namely the
$i$th source CQSO position.  For an error cone $\sigma_r$ the residual
of an event (its contribution to the total $\chi^2$) is $ 2.28  |r -
r^0|^2/\sigma_r^2$. The errors on the QSO positions are negligible in
comparison with those on the UHECRs.  A generalization of Eqn. (\ref{chi2})
could be used if correlations in the errors on the coordinates of a given
UHECR event were available and non-negligible.  

Since there are two degrees of freedom for each event, a residual of
about 2 or less corresponds to a good fit.  The expected fluctuations
in the sum of the residuals is proportinately less than that of any
given residual, so that as $N_U$ increases the statistical power of
the analysis increases. For a given set of UHECR events and associated
hypothetical sources, one determines the Confidence Level of the fit.
The Confidence Level (CL) is the probability, with gaussian
measurement errors, that an ensemble of $N_d = 2 N_U$ measurements
will produce a $\chi^2$ as large or larger than the observed value.
An  explicit formula for determining the CL corresponding to a given
$\chi^2$ and $N_d$ is given in the statistics section of \cite{pdg96}.
For reference,  CL = 0.44 for $\chi^2 = 10.0$ and $N_d = 10$, while if
$\chi^2 = 32.0$ and $N_d = 4$, CL = $2 \times 10^{-6}$ or if $\chi^2 = 
80.0$ and $N_d = 10$, CL = $5 \times 10^{-13}$.  

Table \ref{tab:CTQSO} gives the residuals ($\delta \chi^2$) for each of the
5 events listed in Table \ref{tab:events}, under the hypothesis the
source is the nearest CQSO.  As a check of the method, we make the same
analysis for a second category of ``Test'' QSOs (TQSOs) chosen to have
similar surface density and systematics to the CQSOs, by requiring a QSO
in NED with $0.400 \le z \le 0.600$.  This range of $z$ was intentionally
chosen to include 3C147, the QSO associated with the Fly's Eye event, in
order to mimic the CQSO search.   By using the same portions of the sky,
and considering QSO's rather than another type of object, we avoided introducing
systematic differences between the TQSO and CQSO classes.  There are 7 TQSOs
in the 5 cones of radius 5 deg centered on the 5 UHECR events, giving a
surface density of 0.0178 deg$^{-2}$.  Since there is no physical motivation
that having a redshift in the range $0.4 < z < 0.6$ should be related
to a QSO's acceleration potential, we should NOT find a positive
correlation for the TQSO category.

The first row of Table \ref{tab:CL} gives the probability (CL) to find a
total $\chi^2$ as good as the one observed for CQSOs.  As a check that the
results aren't skewed by having used the properties of 3C147 to define
the CQSO class, we also give the result when the analysis is restricted
to the 4 other events.  Evidently, the hypothesis that UHECR primaries travel
undeflected from compact QSOs provides an excellent explanation for the
observations, and is equally good for the restricted analysis.  The same
is not true for a randomly chosen category of QSO with the same surface
density, as evidenced by the very low confidence level ($< 2.3 \times 10^{-9}$)
for the TQSO fits.  The results for this case are shown in the second row
of Table \ref{tab:CL}.  

By a straightforward Monte Carlo calculation, one can determine the 
probability distribution that {\it randomly distributed} objects having the
same surface density as CQSO's, 0.018 deg$^{-2}$, produce a given value of
$\chi^2$.  The large $\chi^2$ of the TQSOs is in fact typical of the
random-background case:  the probability to find $\chi^2 \ge 61.1$ is
0.59.  The most interesting aspect of the $\chi^2$ probability distribution
is the area below $\chi^2 = 9.02$, since this is the probability 
that the CQSO correlation is a statistical fluctuation.  The results are
given in the bottom line of Table \ref{tab:CL}.  The probability that the
correlation observed betwen CQSOs and UHECRs is accidental is
0.005\footnote{Note that the naive procedure of taking the product of the
probabilities of finding a random source inside each 1-$\sigma$ error region
underestimates this probability by several orders of magnitude due to neglecting
configurations in which some small residuals compensate a large one.}.  Since
the correlation hypothesis is {\it a priori}, there is no reason to restrict
to just 4 events; we include that result in Table \ref{tab:CL} because
the factor-6 increase in the statistical significance of the observed
correlation in going from 4 to 5 events gives an idea of the improvement
in this analysis due to adding a well-measured UHECR event to the sample. 

Let us summarize the underlying assumptions and limitations of the
statistical analysis presented here.  First, we have assumed that the
position errors are gaussian and uncorrelated.  Therefore our results
should be taken as qualitative rather than quantitative indicators of
the relative probabilities.  In the future, cosmic ray experiments
should report as detailed information as possible on the positional
errors of each high energy event.  Secondly, we have assumed that the
density of compact quasars is approximately uniform.  This
may not be valid, either due to some physical structure or to
non-uniformity in the surveys near the different UHECRs, although
there is no obvious reason to suspect this to be the case.

It is important to stress that having an incomplete catalog from which to
choose the best source can only reduce, not exagerate, the quality of the
fit if the alignment hypothesis is correct.  It cannot lead to an incorrect
estimate of the random background probability as long as the surface
density is approximately uniform and is computed from the same
population as used to find the candidate sources.  A dedicated survey within
the 3-sigma error ellipse of each of the UHECR candidates, and also in several
comparable random and nearby patches of sky, would be valuable here.

The candidate sources we have identified should be studied in greater
detail and with better resolution to learn more about their properties
and see if there is a better characterization of the sources. 

The hypothesis that compact radio QSOs are responsible for the highest
energy cosmic rays gets support from a clustering of events noted by
AGASA\cite{akeno:cluster}.  Three of the five UHECR events studied
here have one or two companions -- nearby events with energy near the
GZK bound.  Near 3C147 
we have FE320 at an angular distance of 1.9 deg, Ag62 at 4.0 deg and
the Yakutsk 120-230 EeV event whose error box is large but overlaps. 
Near 0109+224 we have Ag210 at 2.0 deg, HP69 at 4.0 deg and Ag51 at 2.6 deg,
and near 1851+485 
we have Ag110 at 1.5 deg and Ag43 at 1.5 deg. 
The lower energy members of these pairs or triplets either have a low
enough energy to be interpreted as a proton consistent with the GZK
bound and directions consistent with the angular deflection of protons
given in Eqn (\ref{angdev}), or small enough angular distance from the
CQSO source to be interpreted as having been undeflected. 

AGASA has announced\cite{nagano:owl} the observation of another event
with energy above $10^{20}$ eV, but has not yet released its
coordinates or energy.  If the correlation pointed out here is real,
we predict that the new AGASA event will have a compact radio QSO directly
behind it, within measurement errors. A radio search in its vicinity may
be necessary to check this prediction. Since the random probability to
find a CQSO within a 1 deg cone is 0.05, even a few more events 
with good directional information can confirm or cast doubt on the
correlation we have found. HiRes will have an angular accuracy 5 times
better than Fly's Eye and a capacity to collect data which is
approximately an order of magnitude higher. The Auger detector expects
to have a 1 deg resolution, and an even higher data-collection rate.  

To summarize, we have found that the highest energy cosmic rays are
consistent with traveling undeflected from compact radio quasars located
at cosmological distances.  The probablity that this is a statistical
fluctuation is 0.005.  For the moment these results are only a
tantalizing hint that the highest energy cosmic rays may point
directly to their sources and travel cosmological distances.  However
if this hint is born out by future data, nature will have revealed
some new particle physics mechanism involving neutral, GZK-evading
messengers.  Large statistics with improved angular resolution would
allow the properties of the source to be more precisely charcterized.
UHECRs would then complement traditional astronomical tools for
studying these extremely distant and powerful sources and their physics.

\acknowledgments

We gratefully acknowledge correspondence with M. Nagano, P. Sommers and
A. Watson regarding the AGASA, Fly's Eye, and Haverah Park UHECR 
events, and thank P. Kronberg, C. O'Dea and S. Somalwar for helpful
discussions.  Research of GRF supported in part by NSF-PHY-94-32002.


\begin{table}
\begin{center}
\begin{tabular}{|c|c|c|c|c|c|} \hline 
UHECR & date & ENERGY &  RA (deg) & Dec (deg)
      &  $\Delta \Omega$\\ \hline
\hline
FE320 & 15.10.91 & 3.20$^{+ 0.92}_{- 0.94} $ &$ 85.2 \pm 0.5$ & 
   $ 48.0^{+5.2}_{-6.3}$& 2.6 \\ \hline
Ag210 & 03.12.93 & (1.7-2.6) & 18.9  & 21.1 & 8.0 \\ \hline
HP120 &18.04.75& 1.20 $\pm$ 0.10 & 179$ \pm$ 3 &27$ \pm$ 2.8  & 6.7 \\ \hline
Ag110 &06.07.94& 1.10 &280.7 & 48.4& 8.0  \\  \hline
HP105 &12.01.80 & 1.05 $\pm$ 0.08 & $201 \pm 8.7$ & $71 \pm 2.5$ & 7.1
\\ \hline
\end{tabular}
\end{center}
\caption{Events with $E>10^{20} eV$ and solid-angled error less than
10 ${\rm deg}^2$.}  
\label{tab:events}
\end{table}

\begin{table}
\begin{center}
\begin{tabular}{|c|c|c|c|c|c|c|c|} \hline
 \multicolumn{4}{|c|}{Compact QSO} &  \multicolumn{4}{c|}{Test QSO} \\
\cline{1-4} \cline{5-8}
Candidate & z & Sep'n & $\delta \chi^2$ & Candidate & z & Sep'n & $\delta \chi^2$ \\ 
\hline \hline  
3C147 & 0.545 &  111.6 & 1.2 & 3C147 & 0.545 & 111.6 & 1.2 \\ \hline
0109+224 & ... & 119.7 & 3.5 & 0133+207 & 0.425 & 254.4 & 16.3 \\ \hline
1204+281& 2.177 & 138.5 & 0.9 & 1153+317 & 0.418 & 286.3 & 26.4 \\ \hline
1851+485& 1.25 &  89.0 & 2.0 & 1908+483 & 0.513 & 254.9 & 16.2 \\ \hline 
1345+73& 0.29 &  183.4 & 1.4 & 1300+69& 0.570 & 155.1 & 0.9 \\ \hline
\end{tabular}
\end{center}
\caption{Compact and Test QSOs nearest the UHECRs of Table I.
Separation is in arcmin.}      
\label{tab:CTQSO}
\end{table}

\begin{table}
\begin{center}
\begin{tabular}{|c|c|c|c|c|} \hline
Source Class &  $\chi_5^2$ &  Probability &  $\chi_4^2$ &  Probability \\
\hline     \hline
Compact QSO 	& 9.02 	& 0.53   & 7.82 & 0.45  \\ \hline
``Test'' QSO   	& 61.1 	& $2.3 \times 10^{-9}$ 	& 59.9 	& $4.9 \times
10^{-10}$  \\ \hline  
Random QSO   	&  $\le 9.02$ 	& 0.005   & $\le 7.82$  & 0.03 \\ \hline 
\end{tabular}
\end{center}
\caption{Rows 1,2: Probability for Compact and Test QSOs to produce
the observed total $\chi^2$.  Row 3:  Probability for random sources
with surface density of CQSO's (0.018 deg$^{-2}$) to give $\chi^2$
equal or better than observed.  Columns 2,3: all 5 UHECR events; columns
4,5: excluding Fly's Eye Event.}  
\label{tab:CL}
\end{table}

\end{document}